\documentclass[12pt]{JHEP}
\usepackage{amsmath,amsfonts,amssymb}
\usepackage{epsf}

\newcommand{\be}{\begin{equation}}
\newcommand{\ee}{\end{equation}}
\newcommand{\bea}{\begin{eqnarray}}
\newcommand{\eea}{\end{eqnarray}}
\newcommand{\nn}{\nonumber}

\author{Diego H. Correa$^{1,\dagger}$ and Guillermo A.
Silva\thanks{Associated with CONICET} $^{2\ddagger}$ \\
$^1$ Centro de Estudios Cientif\'{\i}cos, Valdivia, Casilla 1469, Chile \\
$^2$ Departamento de F\'\i sica, Universidad Nacional de La Plata
C.C. 67, 1900,\\ La Plata, Argentina
\\
$^\dagger$ \email{dcorrea@cecs.cl}\\
$^\ddagger$ \email{silva@fisica.unlp.edu.ar}}

\title{Dilatation operator and the Super Yang-Mills
duals of open strings on AdS Giant Gravitons}

\abstract{We study the one-loop anomalous dimensions of the Super
Yang-Mills dual operators to open strings ending on AdS giant
gravitons. AdS giant gravitons have no upper bound for their
angular momentum and we represent them by the contraction of
scalar fields, carrying the appropriate R-charge, with a totally
symmetric tensor. We represent the open string motion along AdS
directions by appending to the giant graviton operator a product
of fields including covariant derivatives. We derive a bosonic
lattice Hamiltonian that describes the mixing of these excited AdS
giants operators under the action of the one-loop dilatation
operator of ${\cal N}=4$ SYM. This Hamiltonian captures several
intuitive differences with respect to the case of sphere giant
gravitons. A semiclassical analysis of the Hamiltonian allows us
to give a geometrical interpretation for the labeling used to
describe the fields products appended  to the AdS giant operators.
It also allows us to show evidence for the existence of
continuous bands in the Hamiltonian spectrum.}

\preprint{CECS-PHY-06/20}

\begin{document}

\section{Introduction}
\label{intro}

The most accomplished realization of the AdS/CFT correspondence
conjectures the equivalence between ${\cal N}=4$ Super Yang-Mills
theory in four dimensions and type IIB string theory on the
$AdS_5\times S^5$ background \cite{malda, GKP, Witten}. A great deal
of evidence supporting this equivalence was found after the
realization of a geometrical limit, for strings with large angular
momentum, leading to a plane-wave background where the string theory
can be exactly quantized \cite{Blau, Metsaev}. In this so
called ``BMN" limit, the closed strings energy spectrum was shown to
match the scale dimension spectrum of  certain dual operators
(a.k.a. BMN operators) of the ${\cal N}=4$ gauge theory \cite{BMN}.
The key observation of Berenstein, Maldacena and Nastase \cite{BMN}
was to note that the perturbative contributions to the anomalous
dimensions of BMN operators are suppressed by the square of the
parameter associated with the angular momentum of the dual string.
This fact allowed to engineer a limit in which perturbative
computations on the gauge theory side could be extrapolated to
strong coupling and therefore make a true comparison with string
theory computations beyond the supergravity and BPS approximations.

Another remarkable result in the AdS/CFT large $N$ approximation
($N$ being the rank of the gauge group) was the discovery of
integrable structures governing the anomalous dimensions of ${\cal
N}=4$ SYM operators. Minahan and Zarembo \cite{minahan} showed that
the mixing matrix of anomalous dimensions of single trace operators
of scalar fields, at planar 1-loop approximation, is given by an
integrable SO(6) spin chain Hamiltonian. More recently, the planar
1-loop anomalous dimensions for the full set of local operators
were shown to  be given by the spectrum of an integrable spin chain
Hamiltonian for the full superconformal group \cite{BS}.

Some of these ideas can be extended to the case where one includes
D-branes and open strings. In these cases the spectrum of anomalous
dimensions of the dual operators is given by open spin chain
Hamiltonians. It is possible to work out the boundary conditions for
the open spin chains and determine if they define integrable open
spin chain models.

In \cite{mcgreeve}, non-perturbative BPS states of finite energy on
$AdS_5\times S^5$ were found by considering compact D3-branes
expanded in a $S'^3\subset S^5$. These spherical D-brane solutions,
known as giant gravitons, are supersymmetric and it is also possible
to construct them by wrapping a $S^3\subset AdS_5$ \cite{grisaru,
Hashimoto}. The salient feature of all these configurations is that
the square of the radius of the D-brane is proportional to an
angular momentum $p$ associated with a $S^1\subset S^5$
rigid motion of the brane. This property marks an important distinction between sphere
and AdS giant gravitons: the angular momentum of sphere giants has
an upper bound. On the other hand, giants expanded in AdS have no
bound on their angular momentum. In \cite{BBNS,CJR,david,Btoy}
BPS dual operators to the above giant gravitons were proposed. Quite
interestingly,  some non-BPS perturbations of these operators were
interpreted as open strings attached to giant gravitons in
\cite{BHLN,BBFH,BV,bcv,deM,Agarwal,OY,bcv2}\footnote{Open strings
can also be attached to defect (non-compact) D-branes. In
the dCFT, matter fields serve as boundaries for the spin chain and
define an integrable open spin chain Hamiltonian
\cite{DeWM,OTY}.}. Until recently, the anomalous dimensions of these
non-BPS operators were not studied as extensively as for single
trace BMN operators  discussed in \cite{BMN}. A
computational difficulty being that non-BPS
operators dual to excited giant gravitons have a number of
fields $p$, which taken of order $N$, leads to the failure of
the planar approximation. Nevertheless, it is possible to work out
the combinatorics and extract the leading order result in the large
$N$ limit\footnote{Large $N$ and planar limit are usually taken as synonymous
in the literature.}.

For sphere giant gravitons of maximal size, the resulting boundary
conditions for the one-loop anomalous dimension mixing matrix lead
to an integrable open spin chain \cite{BV}. Computations at
two-loop order have recently shown inconsistency with the Bethe
Ansatz \cite{Agarwal,OY}. So, if the system remains integrable at
higher loops, its integrability will not be implemented by a Bethe
Ansatz. For non-maximal sphere giants, already at one-loop order,
the Hamiltonian is  not solvable by a Bethe Ansatz. However, and
despite the presence of continuous bands in its spectrum, the
Hamiltonian seems to be integrable \cite{bcv2}.

In this work we consider dual  operators to open strings attached to
AdS giant gravitons. The gauge theory description of strings
spinning along the AdS directions of the giant requires to consider
the action of covariant derivatives on the set of scalar operators.
This AdS spinning provides a new parameter $L$ which  will be
crucial for implementing the BMN scaling. We will deal with
operators in a non-compact $sl(2)$ sub-sector  of the superconformal
group. To study the anomalous dimensions spectrum of the set we will
use the complete one-loop dilation operator of ${\cal N}=4$ SYM,
constructed by Beisert in \cite{Beisert1},  and restrict it to the
$sl(2)$ sub-sector. In the present work we will not be concerned on
the integrability of the Hamiltonian giving the one-loop anomalous
dimensions. Instead, we will focus our attention to explore a BMN
limit on both dual descriptions of the system. The paper is
organized as follow. In section \ref{dilatation} we derive the large
$N$ one-loop mixing matrix of anomalous dimensions for the set of
operators dual to open strings attached to AdS giant gravitons. Due
to the variability in the number of sites in the standard spin chain
mapping, we introduce an alternative and more convenient labeling
for the operators as states of a bosonic lattice. We provide a
semiclassical sigma-model action governing the dynamics of the
bosonic Hamiltonian with a large number of sites and also show
evidence of the existence of continuous bands in the spectrum of the
bosonic Hamiltonian. In section \ref{strings} we briefly present the
geometrical description of open strings on AdS giants and guided by
the AdS/CFT correspondence we see how to confer a geometrical
interpretation on the results of section \ref{dilatation}. In
section \ref{discu} we summarize and discuss the results of the
paper. Two appendices \ref{proper} and \ref{ppwave} are devoted to
the relevant combinatorics for the computations of the section
\ref{dilatation}, and the Penrose limit considered in section
\ref{strings}.

\section{Anomalous dimensions for
the duals of open strings on AdS giants}
\label{dilatation}

${\cal N}=4$ SYM operators constructed from a single complex
scalar field $Z$ in the adjoint representation of $U(N)$ are
half-BPS protected.  Polynomial operators in $Z$ have their
R-charge given by the degree of the polynomial and it is possible
to establish a dictionary between these gauge theory half-BPS
operators and half-BPS objects in the dual string theory
\cite{BBNS, CJR,david, Btoy}. For instance, a Schur polynomial in
$Z$, written in the totally symmetric  representation of the
permutations group $S_p$ is identified with an AdS giant graviton. On
the other hand, a Schur polynomial written in the totally
antisymmetric representation is identified with a sphere giant
graviton \cite{BBNS}. Explicitly, the local gauge theory operator
representing an AdS giant graviton with $p$ units of angular
momentum is proposed to be
\begin{equation}
\label{adsgg}
{\cal O}_p
=S^{i_1\cdots i_p}_{j_1\cdots j_p}Z^{j_1}_{\;i_1}\cdots Z^{j_p}_{\;i_p}\,,
\end{equation}
where $S^{i_1\cdots i_p}_{j_1\cdots j_p}$ is a tensor totally symmetric
in all its indices. Its definition together with some useful properties
are displayed in the appendix \ref{proper}.

Open strings attached to giant gravitons (spherical D3-branes)
give rise to non-BPS excitations. These states are mapped in the
gauge theory to operators like (\ref{adsgg}) but with a $Z$ field
replaced by a product of SYM fields and their derivatives,  which,
usually referred to as a word $W$, represents the open string
excitation state. Operators representing excited sphere giant
gravitons were previously studied in
\cite{BHLN,BBFH,BV,bcv,deM,Agarwal,OY,bcv2}. Extending the
analysis to the case of AdS giant graviton operators is
particularly interesting since sphere and AdS giants behave quite
differently when their angular momentum $p$ is increased:
while the angular momentum of a sphere giant is bounded by $N$
\cite{mcgreeve}, there is no upper bound for the angular momentum
of AdS giants (see eqn.(\ref{r2})).

Our attention will be focused on operators of the form,
\begin{equation}
\label{adsggw}
{\cal O}^{W}_p
=S^{i_1\cdots i_p}_{j_1\cdots j_p}
Z^{j_1}_{\;i_1}\cdots Z^{j_{p-1}}_{\;i_{p-1}}W^{j_p}_{\;i_p}\,.
\end{equation}
In particular, we will be interested in open strings following
almost null trajectories in order to be able  to make contact with
gauge theory computations via a BMN limit \cite{BMN}. This
requires to consider open strings spinning fast around the AdS
giant graviton. In gauge theory language, this AdS motion is
represented by the presence of covariant derivatives in the
letters of the word $W$. It will be sufficient for our purposes to
consider letters constructed out from a single scalar field $Z$
and covariant derivatives, taken in a unique spacetime direction,
acting on it. The infinitely many possible letters ${\cal D}^n Z$
can be shown to transform in the infinite dimensional spin
$j=-\frac12$ representation of a non-compact $sl(2)$ subalgebra of
the conformal subalgebra. We will study AdS giant operators
belonging to this $sl(2)$ sub-sector, which can be shown to be
exactly closed \cite{Beisert1}. Therefore, the words $W$ we will
be considering in (\ref{adsggw}) take the form
\begin{equation}
\label{word}
W = Z^{(n_1)}\cdot Z^{(n_2)}\cdots Z^{(n_J)}\,,
\end{equation}
where,
\begin{equation}
\label{word2}
Z^{(n)}= \frac{1}{n!}{\cal D}^n Z\,, \qquad {\cal D}=
{\cal D}_1+i{\cal D}_2\,.
\end{equation}
Here ${\cal D}_1$ and ${\cal D}_2$ are covariant derivatives and the
product in (\ref{word}) should be understood as matrix
multiplication. Our aim is to study the anomalous dimensions
spectrum of the set of operators (\ref{adsggw}) which is conjectured
by AdS/CFT to coincide with the open string excitation spectrum of
the giant graviton. We will calculate the mixing matrix of anomalous
dimension at the large $N$ one-loop approximation.

The complete one-loop dilatation operator $D$ of ${\cal N}=4$ SYM
is known \cite{Beisert1}. We will study its restriction to the AdS
giant operators (\ref{adsggw})-(\ref{word}),
\begin{equation}
\label{dila}
D = D_0 + \frac{g_{\mathrm {YM}}^2N}{8\pi^2}D_1+{\cal O}(g_{YM}^4)\,.
\end{equation}
In (\ref{dila}), $D_0$  gives the classical dimension of the operator,
\begin{equation}
\label{d0} D_0= \sum_{a=0}^\infty (a+1){\rm
tr}Z^{(a)}{\check{Z}}_{(a)}\,,\qquad {\rm with}\quad
\left({{\check{Z}}_{(a)}}\right)^{i}_j = \frac{\delta}{\delta
(Z^{(a)})^j_{\;i}}\,.
\end{equation}
The one-loop contribution $D_1$ in (\ref{dila}) is written as
\begin{eqnarray}
\label{d1}
D_1&=& N^{-1} C_{ab}^{cd}:{\rm tr}[
Z^{(a)}{,\check{Z}}_{(c)}]
[ Z^{(b)}{,\check{Z}}_{(d)}]:\nn\\
&=&  N^{-1} \left(C_{ab}^{cd}+C_{ab}^{dc}\right) :
\left(Z^{(a)}\right)^i_{\,j}
\left({{\check{Z}}_{(c)}}\right)^j_{\,k}
\left(Z^{(b)}\right)^k_{\,l}\left({{\check{Z}}_{(d)}}\right)^l_{\,i}:\nn\\
&&  - N^{-1} \left(C_{ab}^{cd}+C_{ba}^{dc}\right) :
\left(Z^{(a)}\right)^i_{\,j}
\left({{\check{Z}}_{(c)}}\right)^j_{\,k}
\left({{\check{Z}}_{(d)}}\right)^k_{\,l}\left(Z^{(b)}\right)^l_{\,i}:\, ,
\end{eqnarray}
where sums over repeated indices have been omitted and colons
indicate that the variations $\check Z$ do not contract on letters
within the same colons. The coefficients $C_{ab}^{cd}$ can be
obtained from the complete one-loop dilatation operator when its
action is restricted to the $sl(2)$ sub-sector. From eqn. (3.14)
of \cite{Beisert1} one gets
\begin{eqnarray}
 (C_{ab}^{ab}+ C_{ba}^{ba }) &=& -h(a)-h(b)\,,\nn\\
 (C_{a+n\ b-m}^{\;\;\;\;a\;b}+ C_{b-m\ a+n}^{\;\;\;\;b\;a})
 &=&\frac{\delta_{nm}}{|n|}\,,\qquad
 n=-a,\ldots,b\neq 0\,,
 \label{coe}
\end{eqnarray}
where the harmonic numbers $h(a)$ are defined as,
\begin{equation}
\label{hn}
 h(a)=\sum_{i=1}^a\frac{1}{i},\qquad h(0)=0 \,.
\end{equation}
We will now analyze the mixing  of the operators
(\ref{adsggw})-(\ref{word2}) under the action of $D_1$. It will be
helpful to label the operators we will be working with as,
\begin{equation}
\label{go}
{\cal S}(p\ ; a_1,\cdots a_J):=
S^{i_1\cdots i_p}_{j_1\cdots j_p}Z^{j_1}_{\;i_1}\cdots
Z^{j_{p-1}}_{\;i_{p-1}}
(Z^{(a_1)}\cdots Z^{(a_J)})^{j_p}_{\;i_p}\,.
\end{equation}
The rank of the totally symmetric tensor $p$ is
associated with the angular momentum of the giant
 along a $S^1\subset S^5$. The
number of letters $J$ is dual to the angular momentum of the open
string excitation along the same $S^1$ as for $p$, and
the total number of covariant derivatives $L$  distributed in the
word is dual to the open string excitation's angular momentum
along a $S^1\subset AdS_5$ direction. Thus, the integers $a_n$ are
subject to
\begin{equation}
\sum_{n=1}^J a_n = L\, .
\end{equation}
It is necessary to normalize the operators (\ref{go}) so that in the
large $N$ limit, their free correlation functions are of order one.
We define,
\begin{equation}
\label{gon}
{\tilde{\cal S}}(p\ ; a_1,\cdots a_J):=
\sqrt{\frac{(N+1)!}{(N+p-1)!p!(p-1)! N^{J+1}}}
\,{\cal S}(p\ ; a_1,\cdots a_J) \,.
\end{equation}
We will exclude the possibility of $a_1$ or $a_J$ being zero,
since in those cases the identity,
\begin{eqnarray}
S^{i_1 \cdots i_p}_{j_1 \cdots j_p}
Z^{j_1}_{\;i_1} \cdots Z^{j_{p-1}}_{\;i_{p-1}} (Z W)^{j_p}_{i_p}
&=& \frac{1}{p} S^{i_1 \cdots i_{p+1}}_{j_1 \cdots j_{p+1}}
Z^{j_1}_{\;i_1} \cdots
Z^{j_{p}}_{\;i_{p}}W^{j_{p+1}}_{\;i_{p+1}}-\frac{1}{p}
S^{i_1 \cdots i_p}_{j_1 \cdots j_p} Z^{j_1}_{\;i_1} \cdots
Z^{j_p}_{\;i_p} {\rm tr}(W)\,,\nn\\
\label{ide}
\end{eqnarray}
allows to write an operator with a $Z$ at the extreme of a word,
the lhs, in terms of: a bigger giant with a shorter string (first
term, already accounted in the set), and an unexcited D-brane plus
a closed string (second term). When computing the mixing among
operators (see below), the proper account of the normalization
factors for operators shows that the last term in (\ref{ide}) is suppressed by a
factor $1/\sqrt p$ . Since while taking the large $N$ limit we are
taking $p\sim N$, the contribution of the last term in (\ref{ide}) is
irrelevant to the computations.

The action of $D_1$ on operators ${\tilde{\cal S}}(p\ ; a_1,\cdots
a_J)$ is determined by the two derivatives $\check Z$ present in
(\ref{d1}). The possibilities are: (I) both derivatives act on
letters of the word $W$ or (II) one derivative acts on a letter of
the word and the other on one of the $Z$ fields contracted with
the symmetric tensor\footnote{When both derivatives act on $Z$
fields contracted with the symmetric tensor, the contribution is
proportional to $C^{00}_{00}$ which is zero (cf. (\ref{coe})).}.

\vspace{2mm}

\noindent (I) A straightforward computation shows that the leading
order contribution to $D_1$ in the large $N$ limit  comes from the
case where both derivatives act on
consecutive\footnote{Contributions from the action of
$\check Z$ on non-consecutive letters are sub-leading.} letters of
the word $W$. The result is
\begin{eqnarray}
\label{DI}
D_1^{(I)} {\cal S}(p\ ; a_1,\cdots a_J)&=&
-(C_{ab}^{a_1 a_2}+C_{ba}^{a_2 a_1})
\,{\cal S}(p\ ; a,b,a_3,\cdots a_J)\\
&&-(C_{ab}^{a_2 a_3}+C_{ba}^{a_3 a_2})
\,{\cal S}(p\ ; a_1,a,b,\cdots a_J)-\cdots\nn\\
&&-(C_{ab}^{a_{J-1} a_J}+C_{ba}^{a_J a_{J-1}})
\,{\cal S}(p\ ; a_1,\cdots a_{J-2},a,b)+
{\cal O}(\frac1N)\,.\nn
\end{eqnarray}
Using the identity (\ref{ide}) we rewrite apart the cases $a=0$ in
the first line and $b=0$ in the last line. Taking into account the
normalization of the operators (\ref{gon}) we obtain,
\begin{eqnarray}
\label{DI2}
D_1^{(I)} {\tilde{\cal S}}(p\ ; a_1,\cdots a_J)&=&
-(C_{a' b}^{a_1 a_2}+C_{ba'}^{a_2 a_1})\,
{\tilde{\cal S}}(p\ ;a',b,a_3,\cdots a_J)\nn\\
&&-(C_{ab}^{a_2 a_3}+C_{ba}^{a_3 a_2})\,
{\tilde{\cal S}}(p\ ;a_1,a,b,\cdots a_J)-\cdots\nn\\
&&-(C_{ab'}^{a_{J-1}a_J}+C_{b'a}^{a_J a_{J-1}})\,
{\tilde{\cal S}}(p\ ; a_1,\cdots a_{J-2},a,b')\nn\\
&&-\sqrt{1+\frac{p}{N}}(C_{0b}^{a_1 a_2}+C_{b0}^{a_2 a_1})\,
{\tilde{\cal S}}(p+1;b,a_3,\cdots a_J)\nn\\
&&-\sqrt{1+\frac{p}{N}}(C_{a0}^{a_{J-1} a_J}+C_{0a}^{a_J a_{J-1}})\,
{\tilde{\cal S}}(p+1 ;a_1,\cdots a_{J-2},a)\nn\\
&&+{\cal O}(\frac1{\sqrt{p}})\,.
\end{eqnarray}
Here a primed repeated index indicates that its summation
excludes the value zero. The last two terms in (\ref{DI2}) show
that the dilatation operator mixes states with words of different
lengths. This instance is similar to that of non-maximal sphere
giant gravitons \cite{bcv,bcv2} and the mixing between operators
with words of different lengths could have been expected. The
variation in the number of letters of the word $W$ can be
pictured, from the string point of view, as coming from the
exchange of the angular momentum along the $S^1\subset S^5$ between the
string and the giant. The open string gets dragged by the movement
of the giant graviton while propagating. Notice also that the
factor describing the mixing of words of different lengths is
$\sqrt{1+p/N}$. The sign inside the square root has changed with
respect to the similar factor appearing in the sphere giant case.
This last fact reflects that $p$ can increase arbitrarily for AdS
giants.

\vspace{2mm}

\noindent(II) The leading order result, considering as before
$p\sim N$, is
\begin{eqnarray}
\label{DII}
D_1^{(II)} {\tilde{\cal S}}(p\ ; a_1,\cdots a_J)&=&
-\left(1+\frac p N\right)
(C_{b 0}^{a_1 0}+C_{0 b}^{0 a_1})\,
{\tilde{\cal S}}(p\ ; b,a_2,\cdots a_J)\nn\\
&&-\left(1+\frac p N\right)
(C_{b 0}^{a_J 0}+C_{0 b}^{0 a_J})\,
{\tilde{\cal S}}(p\ ; a_1,\cdots a_{J-1},b)\nn\\
&&-\sqrt{1+\frac{p}{N}}
(C_{a'b}^{0 a_1}+C_{b a'}^{a_1 0})\,
{\tilde{\cal S}}(p-1;a',b,a_2,\cdots a_J)\nn\\
&&-\sqrt{1+\frac{p}{N}}
(C_{a b'}^{a_{J} 0}+C_{b' a}^{0 a_J})\,
{\tilde{\cal S}}(p-1 ; a_1,\cdots a_{J-1},a,b')\nn\\
&&-\frac{p}{N}(C_{0b}^{0 a_1}+ C_{b 0}^{a_1
0}+C_{0b}^{a_1 0}+ C_{b 0}^{0 a_1})\,
{\tilde{\cal S}}(p\ ;b, a_2,\cdots a_J)\nn\\
&&-\frac{p}{N}(C_{0b}^{0 a_J}+ C_{b 0}^{a_J
0}+C_{0b}^{a_1 0}+ C_{b 0}^{0 a_1})\, {\tilde{\cal S}}(p\
;a_1,\cdots a_{J-1},b)\nn\\&&+ {\cal O}(\frac1{\sqrt{p}})\,.
\end{eqnarray}
Defining the parameter
\begin{equation}
\alpha\equiv\sqrt{1+\frac p N}\,.
\label{alfa}
\end{equation}
and using the coefficients (\ref{coe}), the result (\ref{DI2}) can
be rephrased as
\begin{eqnarray}
\label{DI3}
D_1^{(I)} {\tilde{\cal S}}(p\ ; a_1,\cdots a_J)&=&
(h(a_1)+h(a_2))\,{\tilde{\cal S}}(p\ ; a_1,\cdots a_J)\nn\\
&&-\sum_{n=-a_2}^{a_1-1}\frac{1}{|n|}\,
{\tilde{\cal S}}(p\ ; a_1-n,a_2+n,,\cdots a_J)\nn\\
&&+(h(a_2)+h(a_3))\,{\tilde{\cal S}}(p\ ; a_1,\cdots a_J)\nn\\
&&-\sum_{n=-a_3}^{a_2}\frac{1}{|n|}\,
{\tilde{\cal S}}(p\ ; a_1,a_2-n,a_3+n,\cdots a_J)-\cdots\nn\\
&&+(h(a_{J-1})+h(a_{J}))\,{\tilde{\cal S}}(p\ ; a_1,\cdots, a_J)\nn\\
&&-\sum_{n=-a_{J}}^{a_{J-1}-1}\frac{1}{|n|}\,
{\tilde{\cal S}}(p\ ;a_1,\cdots,a_{J-1}-n, a_J+n)\nn\\
&&- \frac{\alpha}{a_1}\,{\tilde{\cal S}}(p+1;a_1+a_2,a_3,\cdots a_J)\nn\\
&&- \frac{\alpha}{a_J}\,{\tilde{\cal S}}(p+1;a_1,\cdots,a_{J-1}+
a_J) +{\cal O}(\frac1{\sqrt{p}})\,.
\end{eqnarray}
The first six lines can be identified with the action of an open
$sl(2)$ spin chain Hamiltonian  under the standard
identification $Word \leftrightarrow Spin~Chain~State$
\cite{minahan,BS,Beisert1}. The last two lines indicate that sites
can be annihilated at the boundaries of the chain. Similarly,
(\ref{DII}) can be rewritten as,
\begin{eqnarray}
\label{DII2}
D_1^{(II)} {\tilde{\cal S}}(p\ ; a_1,\cdots a_J)&=&
\alpha^2(h(a_1)+h(a_J))\,{\tilde{\cal S}}(p\ ; a_1,\cdots a_J)\nn\\
&&- \alpha \sum_{n=1}^{a_1}\frac{1}{n}
{\tilde{\cal S}}(p-1;n,a_1-n,a_2,\cdots a_J)\nn\\
&&- \alpha \sum_{n=1}^{a_J}\frac{1}{n}
{\tilde{\cal S}}(p-1;a_1,\cdots,a_{J-1},a_J-n,n) \nn\\
&&- (\alpha^2-1)(h(a_1-1)+h(a_J-1))\,
{\tilde{\cal S}}(p\ ; a_1,\cdots a_J)\nn\\
&&+{\cal O}(\frac1{\sqrt{p}})\,.
\end{eqnarray}
The first and last lines represent the action of identity terms,
while the two middle ones show that sites can  be created at the
boundaries of the chain.

\vspace{2mm} Describing different words as open $sl(2)$ spin chain
states is not the most convenient picture. Since sites can be
created or annihilated at the boundaries, one would have to deal
with spin chains of variable length. In order to  find a more
appropriate labeling of the  words set, it is crucial to note that
the total number of covariant derivatives in the word $W$ is
conserved under the action of the 1-loop dilatation operator. We
choose then to label the words (\ref{word}) by stating the number
of $Z$ fields between consecutive covariant derivatives. Consider,
for instance, the word
\begin{equation}
Z^{(1)}Z^{(0)}Z^{(2)}Z^{(1)}
\sim {\cal D}Z Z {\cal D}{\cal D}Z {\cal D}Z\,.
\end{equation}
We specify it by indicating that there are two $Z$s between the
first and the second  derivative ${\cal D}$, no one between the
second and the third and one $Z$ between the third and the fourth
derivative. A word with $L+1$ covariant derivatives will then be
labeled as a  bosonic state of a lattice with $L$ sites,
\begin{equation}
{\cal D}Z^{n_1}{\cal D}Z^{n_2}{\cal D}\cdots {\cal D}Z^{n_L}{\cal D}Z
\qquad\leftrightarrow\qquad |n_1,n_2,\cdots,n_L\rangle\,,
\label{L}
\end{equation}
with $n_i = 0,1,\ldots$.  The variability of the $sl(2)$ spin
chain length is translated, in the labeling (\ref{L}), into a
variability of the total occupation number of the lattice. This
variability will take place at the size of the giant graviton
expense. However, since the probability of a $Z$ entering or
leaving the word is the same, we expect that $p$ can be
consistently taken as a constant. A posteriori we will check that
occupation number of $Z$s for the ground state is much smaller
than $p$. Summarizing, the total number of bosons in the bosonic
lattice is equal to the total number of $Z$s in the word.

To translate the action of $D_1$ (\ref{DI3})-(\ref{DII2}) to the
bosonic language (\ref{L}), we introduce shift operators
$\hat{a}_i^\dagger$ and $\hat{a}_i$ that rise and lower the
occupation number of $i^{\rm th}$ site
\begin{equation}
\label{ca1}
\hat{a}_i^\dagger |n_i\rangle = |n_i+1\rangle\,,\qquad
\hat{a}_i |n_i\rangle = |n_i-1\rangle\,.
\end{equation}
Note that their action does not involve the square roots of the
standard oscillator-like operators, therefore,
\begin{equation}
\label{ca2}
\hat{a}_i \hat{a}_i^\dagger = I\,,\qquad
\hat{a}_i^\dagger \hat{a}_i =I -P^0_i \equiv I-|0\rangle\langle 0|_i\,.
\end{equation}
Consider the word that begins as
\begin{equation}
\begin{array}{ccccc}
& a_1 &  & a_2 &  \\
Z^{(a_1)}Z^{(a_2)} \cdots \,\sim &\overbrace{{\cal D}\cdots{\cal D}}&Z&
\overbrace{{\cal D}\cdots{\cal D}} & Z\cdots \,.\\
\end{array}
\end{equation}
The only restriction we have is that $a_1\neq0$ (cf. (\ref{ide})).
The first $a_1-1$ sites are empty states in the bosonic language
and the $a_1^{~\rm th}$ is necessarily occupied. The amount of
bosons occupying it depends on the subsequent $a_i$,
\begin{equation}
Z^{(a_1)}Z^{(a_2)} \cdots \quad \leftrightarrow\quad
\left\{
\begin{array}{cccl}
a_1-1 &  &  &  \\
|\overbrace{0,\cdots,0}, & 1, &\cdots \rangle&
\quad {\rm if\ } a_2\neq 0\,, \\
 a_1-1 &  & &  \\
|\overbrace{0,\cdots,0}, & 2, &\cdots \rangle  &
\quad {\rm if\ }a_2 = 0 \wedge\,  a_3\neq 0 \,,\\
 \vdots &  & &  \\
 \vdots
\end{array}
\label{label}
\right.
\end{equation}
Let us consider in detail the translation of some  of the terms in
(\ref{DI3}) and finally present the complete result. We can think
of $D_1$ as the Hamiltonian of a bosonic lattice. Defining $H
=\lambda D_1$, the diagonal terms in (\ref{DI3}) can be understood
as amounts of energy for each site occupied.  Their total
contribution is
\begin{equation}
E_{osc} \sim \lambda(h(a_1)+ 2h(a_2)+\cdots+2h(a_{J-1})+h(a_J))\,,
\end{equation}
where we have introduced the 't Hooft parameter
$\lambda=\frac{g_{YM}^2 N}{8\pi^2}$. This total contribution can
be obtained in the following way: for each site of the lattice
there is no contribution if it is empty, if the site is occupied
the amount is independent of the number of bosons in the
site\footnote{This is the reason for having chosen the shift
operators (\ref{ca1}) instead of ordinary creation and
annihilation operators}. However, the amount for each occupied
site depends on the occupancy of their neighbors. The contribution
in question is $h(e_L+1)+h(e_R+1)$, where $e_L$ is the number of
consecutive empty sites (if any) to the left and $e_R$ is the
number of consecutive empty sites (if any) to the right. These
diagonal terms of the lattice Hamiltonian are written as
\begin{eqnarray}
H_{osc}&=&\lambda\sum_{m=1}^{L-1}\sum_{l=1}^{L-m}
\frac{1}{m}\ ({\hat a}_{l}^\dagger{\hat a}_l
+{\hat a}_{l+m}^\dagger{\hat a}_{l+m})
\left(\prod_{s=l+1}^{l+m-1} P^0_s\right) \nn\\
&& + \lambda\sum_{l=1}^{L}\frac{1}{l}\
{\hat a}_{l}^\dagger {\hat a}_l \left(\prod_{s=1}^{l-1} P^0_s\right)+
\sum_{l=1}^{L}\frac{1}{L+1-l} {\hat a}_{l}^\dagger {\hat a}_l
\left(\prod_{s=l+1}^{L} P^0_s\right)\,.
\end{eqnarray}
There are also terms in (\ref{DI3}) that represent the exchange of
$n$ covariant derivatives between consecutive letters of the word.
In the bosonic labeling, they are seen as hopping terms. Since more
than one covariant derivative can be exchanged, the hopping is not
only between nearest neighbors. A boson can be exchanged between
non-nearest neighbor sites as long as all the sites between them are
empty. These hopping terms can be written as,
\begin{eqnarray}
H_{hopping}= -\lambda\sum_{m=1}^{L-1}\sum_{l=1}^{L-m}
\frac{1}{m}\ ({\hat a}_{l+m}^\dagger {\hat a}_l
+{\hat a}_{l}^\dagger {\hat a}_{l+m})
\left(\prod_{s=l+1}^{l+m-1} P^0_s\right) \,.
\end{eqnarray}
A similar analysis can be repeated to rephrase the terms
corresponding to the creation and annihilation of letters at the
boundaries as well as the diagonal terms in (\ref{DII2}).
Altogether, the action of $D_1$ on the operators corresponding to
open strings on AdS giant gravitons, is given by the action of the
following bosonic lattice Hamiltonian,
\begin{eqnarray}
\label{hamiltonian}
H &=&
\lambda\sum_{m=1}^{L-1}\sum_{l=1}^{L-m}
\frac{1}{m}({\hat a}_{l}^\dagger -
{\hat a}_{l+m}^\dagger)( {\hat a}_l- {\hat a}_{l+m})
\left(\prod_{s=l+1}^{l+m-1} P^0_s\right) \\
&& + \lambda\sum_{l=1}^{L}\frac{1}{l}
\left({\hat a}_{l}^\dagger {\hat a}_l +\alpha^2
- \alpha ({\hat a}_l + {\hat a}_{l}^\dagger )
+(1-\alpha^2)P^0_l\right)\left(\prod_{s=1}^{l-1} P^0_s\right)\nn\\
&&+\lambda\sum_{l=1}^{L}\frac{1}{L+1-l}
\left({\hat a}_{l}^\dagger {\hat a}_l +\alpha^2
- \alpha ({\hat a}_l + {\hat a}_{l}^\dagger )
+(1-\alpha^2)P^0_l\right) \left(\prod_{s=l+1}^{L} P^0_s\right)\,,\nn
\end{eqnarray}
The terms proportional to $\alpha$ in the second and third lines
of (\ref{hamiltonian})  represent sinks and sources for bosons.
Bosons can be created or annihilated at the $l^{\rm th}$ site of
the lattice, as long as all sites in between the $l^{\rm th}$ site
and one of the boundaries are empty. As a consequence, the total
number of bosons does not commute with the Hamiltonian. This
represents a serious difficulty in trying to diagonalize
(\ref{hamiltonian}): their eigenvalues will not have a definite
number of bosons. Nevertheless, it is possible to  build a uniform
coherent eigenstate with zero eigenvalue. This is nothing but the
ground state of (\ref{hamiltonian}),
\begin{eqnarray}
\label{ground} |\Psi_0\rangle =
\left|\alpha^{-1},\ldots, \alpha^{-1}\right\rangle
=\left(\frac{\alpha^2-1}{\alpha^2}\right)^{L/2}\!\!\!\!\!\sum_{n_1,
\ldots, n_L = 0}^\infty \!\!\!\!\! {\alpha^{-(n_1 +
\cdots + n_L)}}|n_1,\ldots, n_L\rangle\;.
\end{eqnarray}
However, we do not know how to solve the eigenvalue problem  in
general.

\subsection{The Semiclassical Limit}

We will show in this section that the dynamics, in the $L
\rightarrow \infty$ limit, given by the Hamiltonian
(\ref{hamiltonian}) is governed by a semiclassical sigma-model
action. The action will be obtained by taking a continuum limit of
the path integral representation of the evolution operator written
 in a coherent states basis.  It was shown in \cite{kru}
that the semiclassical action obtained for the Heisenberg spin
chain Hamiltonian, which describes the planar 1-loop anomalous
dimension of single trace operators in a $su(2)$ sub-sector, can
be related to the closed string action when appropriate gauge
choice and specific limits are taken. A similar identification was
done for single trace operators in a $sl(2)$ sub-sector in
\cite{Stefanski, Bel,Bel2}.

The semiclassical action for the dynamics in the  $L\rightarrow
\infty$ limit reads,
\begin{eqnarray}
\label{saction} S \!&=&\! \int dt \left(i \langle z_1\ldots z_L
|\frac{d}{dt}|z_1\ldots z_L\rangle - \langle z_1\ldots z_L |H
|z_1\ldots z_L\rangle \right)\,,
\end{eqnarray}
We are dealing with a bosonic Hamiltonian and the coherent states
we use are defined to be eigenstates of the shift operator $\hat
a$,
\begin{equation}
\label{coherent}
|z\rangle =\sqrt{1-|z|^2}\
\sum_{n=0}^{\infty} z^n|n\rangle\,,  \qquad
{\rm with}\;\;\;|z|<1\; ,
\end{equation}
As known, coherent states constitute a non-orthogonal and
overcomplete basis. The overlapping between states is
\begin{equation}
\langle z| z'\rangle =
\frac{\sqrt{1-|z|^2}\sqrt{1-|z'|^2}}{1-\bar z z'}\;.
\end{equation}
We parameterize the coherent  states as $z_l(t)=
u_l(t)e^{i\phi_l(t)}$. The first term in the integrand o
(\ref{saction}) takes then the form
\begin{eqnarray}
\label{deri1}
i \langle z_1\ldots z_L |\frac{d}{dt}|z_1\ldots z_L\rangle
&=& -\sum_{l=1}^{L} \frac{u_l^2 \dot \phi_l}{1-u_l^2}\,.
\end{eqnarray}
In the limit of large $L$, the functions $u_l(t)$ and $\phi_l(t)$
can be considered  to be continuous functions $u(t,\sigma)$ and
$\phi(t,\sigma)$ with $0\leq\sigma\leq 1$, the sum over $l$ being
converted into an integral over $\sigma$,
\begin{eqnarray}
\label{deri2}
i \langle z_1\ldots z_L |\frac{d}{dt}|z_1\ldots z_L\rangle
&=& -L \int_0^1 d\sigma\, \frac{u^2 \dot \phi}{1-u^2}\,.
\end{eqnarray}
Here a dot denotes a derivative with respect to $t$.  For the
second term in the integrand of (\ref{saction}) one obtains
\begin{eqnarray}
 \langle H\rangle &=& \langle z_1\ldots z_L|
H|z_1\ldots z_L\rangle \nn
\\
&=& \lambda\sum_{m=1}^{L-1} \sum_{l=1}^{L-m}\frac{1}{m}
(u_{l+m}-u_l)^2
\left(\prod_{s=l+1}^{l+m-1}(1-u_s^2)\right)\nn
\\
&&-\lambda\sum_{m=1}^{L-1}
\sum_{l=1}^{L-m}\frac{1}{m} 2 u_l u_{l+m}(\cos(\phi_{l+m}-\phi_l)-1)
\left(\prod_{s=l+1}^{l+m-1}(1-u_s^2)\right)
\nn
\\
&& +\lambda\sum_{m=1}^{L} \frac{1}{m}\!
\left(\alpha^2 u_m^2+1-2 \alpha u_m \cos\phi_m \right)
\left(\prod_{s=1}^{m-1}(1-u_s^2)\right)\nn\\
&&+\lambda\sum_{m=1}^{L} \frac{1}{L+1-m}\!
\left(\alpha^2 u_m^2+1-2 \alpha u_m \cos\phi_m \right)
\left(\prod_{s=m+1}^{L}(1-u_s^2)\right)\,.\label{hami1}
\end{eqnarray}
In the large $L$ limit of (\ref{hami1}), sums over $l$ (in the
first two lines) are going to be approximated by integrals over a
continuous variable $\sigma$, while sums over $m$ are going to be
approximated by geometric sums. Identifying $f_l \simeq
f(\sigma)$,
\begin{equation}
f_{l+m} \simeq f(\sigma)+\frac{m}{L}f'(\sigma)
+\frac{1}{2}\left(\frac{m}{L}\right)^2f''(\sigma)+
{\cal O}(m^3/L^3)\,.
\end{equation}
Approximating the sums as indicated, the contributions to
(\ref{hami1}) can be gathered order by order in powers of $1/L$.
The result is
\begin{eqnarray}
\label{hami2} \langle H\rangle \!&=&\! -\frac{\lambda}{L}\int_0^1d\sigma\,
\frac{1}{u^4}\left(u'^2+u^2\phi'^2\right)
\\
&& + \lambda\sum_{m=1}^{L} \left.\frac{{(1-u^2)}^{m-1}}{m}
\left(\alpha^2 u^2+1-2 \alpha u \cos\phi \right)\right|_{\sigma=0}\nn
\\
&&+ \frac{\lambda}{L}\sum_{m=1}^{L} \frac{m-1}{m}(1-u^2)^{m-1}
\left(2u'(u-\alpha\cos\phi)+2\alpha u \sin\phi \phi'
\vphantom{\frac12}\right.\nn\\
&&\left.\left.-(m-2)\frac{u u'}{1-u^2}(u^2+\alpha^2-2\alpha u\cos\phi)
-m(1-\alpha^2)u u' \right)\right|_{\sigma=0}\nn
\\
&& + \lambda\sum_{m=1}^{L} \left.\frac{{(1-u^2)}^{m-1}}{m}
\left(\alpha^2 u^2+1-2\alpha u\cos\phi\right)\right|_{\sigma=1}\nn
\\
&& + \frac{\lambda}{L}\sum_{m=1}^{L} \frac{m-1}{m}(1-u^2)^{m-1}
\left(2u'(u-\alpha\cos\phi)+2\alpha u \sin\phi \phi'
\vphantom{\frac12}\right.\nn\\
&&\left.\left.-(m-2)\frac{u u'}{1-u^2}(u^2+\alpha^2-2\alpha u\cos\phi)
-m(1-\alpha^2)u u' \right)\right|_{\sigma=1}\!\!\!\!+{\cal O}(1/L^2)\,.\nn
\end{eqnarray}
Primes denote derivatives with respect to $\sigma$. The bulk term
of this semiclassical Hamiltonian is of order $\lambda/L$.
However, the boundary terms leading order is $\lambda$. Then, the
semiclassical configurations with the lowest anomalous dimension
are going to be those satisfying the following Dirichlet boundary
conditions,
\begin{equation}
\left.\left(\alpha^2 u^2+1-2\alpha u\cos\phi\right)
\right|_{\sigma=0,1}=0 \,.
\end{equation}
This is equivalent to require,
\begin{eqnarray}
u|_{\sigma = 0, 1} &=& \frac{1}{\alpha} \;, \nn
\\
\phi|_{\sigma = 0, 1} &=& 0\;.
\label{bc}
\end{eqnarray}
Note that imposing the boundary conditions (\ref{bc}) cancels not
only the boundary terms of order $\lambda$ but also those of order
$\lambda/L$. Finally, the semiclassical action for the large $L$
limit with boundary conditions (\ref{bc}) is,
\begin{eqnarray}
\label{sa}
S\!&=&\! -L \int dt \int_0^1 d\sigma\,
\left(\frac{u^2 \dot \phi}{1-u^2}
-\frac{\lambda}{L^2 u^4}\left(u'^2+u^2\phi'^2\right)\right)\,.
\end{eqnarray}
At the end of section \ref{som}, we will give a geometrical
interpretation of the coherent states parameters $u$ and $\phi$.
To this end, it is convenient to introduce the variable
$r\equiv1/u$, taking values $r\in(1,\infty)$. The action
(\ref{sa}) is rewritten as
\begin{eqnarray}
\label{sa2} S
\!&=&\! -L \int dt \int_0^1 d\sigma\,
\left(\frac{\dot \phi}{r^2-1}
-\frac{\lambda}{L^2}\left(r'^2+r^2\phi'^2\right)\right)\,,
\end{eqnarray}
and the Dirichlet boundary conditions (\ref{bc}) as,
\begin{eqnarray}
\label{bcr}
r|_{\sigma = 0, 1} &=& {\alpha}=\sqrt{1+\frac pN} \,,
\\
\phi|_{\sigma = 0, 1} &=& 0\,.
\label{bcfi}
\end{eqnarray}
The ground state of this semiclassical action is the constant
configuration $r=\alpha$ and $\phi=0$, which is a coherent state
$z=1/\alpha$ uniformly distributed along the lattice. This is
precisely the ground state (\ref{ground}).

\subsection{Continuous bands}
\label{bands}

In this section we argue that the Hamiltonian (\ref{hamiltonian})
has continuous bands in its spectrum. The variability in the mean
occupation number together with these continuous bands indicate
that it is possible to construct states whose evolution gives a
growing mean occupation number.

Among the family of Hamiltonians (\ref{hamiltonian}) we are only
able to exactly diagonalize the case $L=1$.  The one site, the
Hamiltonian is,
\begin{eqnarray}
H &=&2\lambda\left(\hat a^\dagger\hat a +\alpha^2 -
\alpha (\hat a + \hat a^\dagger )
+(1-\alpha^2)P^0)\right)\nn\\
&=&2\lambda\alpha^2\left(\hat a^\dagger \hat a + \frac1{\alpha^2}
- \frac1{\alpha} (\hat a + \hat a^\dagger )\right)\,.
\end{eqnarray}
A ground state and a continuum after a gap constitute the
whole set of eigenstates,
\begin{eqnarray}
|\Psi_0\rangle &=&\left(\frac{\alpha^2-1}{\alpha^2}\right)^{1/2}
\sum_{n= 0}^\infty \frac{1}{\alpha^{n}}|n\rangle\,,\\
|\Psi(k)\rangle  &=& \sum_{n=0}^{\infty}\left(\sin kn -
\frac1\alpha\sin k(n+1)\right)|n\rangle\,,\quad{\rm with\ } 0\leq
k\leq\pi\,.
\end{eqnarray}
The energy of the ground state is zero, while the energy of the
states in the band is
\begin{equation}
E(k)= 2\lambda(1 -2 \alpha\cos k+\alpha^2)\,.
\label{b1}
\end{equation}
The gap between the ground state and the band is
\begin{equation}
2\lambda(\alpha-1)^2\,.
\label{gap}
\end{equation}
States in the band satisfy a delta-function normalization.
Normalizable wave-packets can be built out of them, whose mean
occupation numbers grow monotonically as evolution takes place.

To analyze the existence of continuous bands for larger $L$
we consider the system semiclassically. As we have already argued, the action
(\ref{saction}) governs the dynamics of our system in the
$L\to\infty$ limit. The idea we pursue is simple: energies for
which orbits in phase space are open, correspond to a continuum in
the spectrum of the quantized system.

To begin with, notice that conjugate momenta $p_{u_l}$ and
$p_{\phi_l}$ are subject to constraints\footnote{From
(\ref{deri1}) one gets $p_{u_l}= 0$ and
$p_{\phi_l}=\frac{u_l^2}{u_l^2-1}$.}. Therefore, the phase space
coincides with the configuration space $(u_l,\phi_l)$, which is a
product of $L$ discs of radii r=1. We are interested in the values
of the semiclassical Hamiltonian (\ref{hami1}), which allow open
orbits in the phase space. Since the energy is a conserved
quantity, orbits are constrained to hypersurfaces of constant
energy. Therefore, open orbits are possible for energies, whose
hypersurfaces intersect the boundary of the phase space. We are be
only interested in the minimal amount of energy that a state would
need to reach a continuous band. Then, we have to compute the
minimum of each of the functions resulting when one of the $u_l$
takes the values 1. These $L$ functions are bounded from below by
polynomials on $u_l$
\begin{eqnarray}
\label{ML}
\left.\langle H\rangle\right|_{u_k=1} \geq
{\cal M}_L^k
&=& \lambda\sum_{m=1}^{L-1} \sum_{l=1}^{L-m}\frac{1}{m}
(u_{l+m}-u_l)^2
\left.\left(\prod_{s=l+1}^{l+m-1}(1-u_s^2)\right)\right|_{u_k=1}
\\
&& +\lambda\sum_{m=1}^{L} \frac{1}{m}\!
\left.\left(\alpha^2 u_m^2+1-2 \alpha u_m\right)
\left(\prod_{s=1}^{m-1}(1-u_s^2)\right)\right|_{u_k=1}\nn\\
&&+\lambda\sum_{m=1}^{L} \frac{1}{L+1-m}\!
\left.\left(\alpha^2 u_m^2+1-2 \alpha u_m \right)
\left(\prod_{s=m+1}^{L}(1-u_s^2)\right)\right|_{u_k=1}\,.\nn
\end{eqnarray}
which are obtained by setting all angles $\phi_l$ to zero. The
minima of the ${\cal M}_L^k$ then give us the minima of
$\left.\langle H\rangle\right|_{u_k=1}$. It is straightforward to
verify that the stationary point equations are satisfied with the
following values,
\begin{eqnarray}
\frac{1}{u_l^*} &=& \left(\frac{1-\alpha}{k}\right)l+\alpha
\qquad \qquad\qquad\qquad{\rm if}\; l \leq k\ , \nn
\\
\frac{1}{u_l^*}&=&\left(
\frac{\alpha-1}{(L+1-k)}\right)l+\frac{L+1 -k\alpha}{(L+1-k)}
\qquad {\rm if}\; l > k\,.
\end{eqnarray}
One can also see that these critical points are  minima of each
${\cal M}_L^k$, and the energy evaluated on them is
\begin{equation}
\left.\langle H\rangle\right|_{u_l=u_l^*,\phi_l=0}
=\frac{\lambda(L+1)(\alpha-1)^2}{k(L+1-k)}\,.
\label{ex}
\end{equation}
It is not difficult to see that the intersection where the central
site takes the value one gives a minimum of the expression
(\ref{ex}). Summarizing, the minimal amount of energy for having
open orbits is
\begin{equation}
E_{\rm cont}= \left\{
\begin{array}{cll}
\frac{4\lambda(\alpha-1)^2}{(L+1)}& &{\rm if}\ L\ {\rm is\ odd}\,, \\
\frac{4\lambda(L+1)(\alpha-1)^2}{L(L+2)}& &{\rm if}\ L\ {\rm is\ even}\,.
\end{array}
\right.
\end{equation}
This analysis enables us to conclude that in the large $L$ limit,
the minimal energy to reach the lowest continuous band in the
spectrum is,
\begin{equation}
E_{\rm cont}= \frac{4\lambda(\alpha-1)^2}{L}\,.
\end{equation}
Operators dual to long semiclassical strings, with energy order
$\lambda/L$, could show a growing occupation number. The existence
of these configurations indicate that the D-brane might be
unstable if it is excited with long strings.

It is possible to give a picture of these configurations from the
geometrical viewpoint. The motion of the giant along the $\psi$
direction of the $S^5$  exerts centrifugal forces on long strings
attached to it. For sufficiently long strings, these forces could
excess the string tension. To see if this phenomenon constitutes a
D-brane instability, it is necessary to study how the diminution in
giant size back-react on these configurations with growing
occupation number. However, this analysis is beyond the validity of
the approximations we made in the derivation of (\ref{hamiltonian}).

\section{Open strings on AdS giant gravitons}
\label{strings}

Let us briefly state some of the geometrical properties of open
strings ending on AdS giant gravitons. We will see that some of
them have already appeared as outcomes of the Hamiltonian
(\ref{hamiltonian}).

We write the metric of $AdS_5\times S^5$ in global coordinates
\begin{equation}
\label{global}
ds^2 = R^2(-\cosh ^2\!\! \rho\, dt^2 +d\rho^2
+\sinh^2\!\!\rho\, d{\Omega_3}^2  +
d\theta^2 + \cos ^2\! \theta\, d\psi^2 +
\sin^2\!\theta\, d{\Omega'_3}^{2})\, ,
\end{equation}
and the 3-sphere metrics as
\begin{eqnarray}
d{\Omega_3}^{2} &=& d\varphi^2 + \cos^2\!\varphi\, d\eta^2
+\sin^2\!\varphi\, d\xi^2\, ,\nn\\
d{\Omega'_3}^{2} &=& d{\varphi'}^2 + \cos^2\!{\varphi'} d{\eta' }^2
+\sin^2\!{\varphi'} d{\xi'}^2\, .
\label{omegas}
\end{eqnarray}
A RR 4-form potential supports the geometry, its self-dual field
strength possessing $N$ units of flux on the $S^5$. The
supergravity equations of motion relate the radius $R$ in
(\ref{global}) to the $F_5$ flux on $S^5$ according to $R^4 = 4\pi g_s N
\alpha'^2$.

Giant gravitons are spinning spheric D3-branes that expand in
either the 3-spheres ${\Omega'}_3$ or ${\Omega}_3$. They spin
rigidly along the $\psi$ direction inside the $S^5$ and are
located either at $\rho=0$ and $\theta=\theta_0$, or at $\theta=0$
and $\rho=\rho_0$ \cite{mcgreeve, grisaru, Hashimoto}. The former
correspond to D-branes expanded into a $S'^3\subset S^5$ and we
refer to them as sphere giant gravitons. The latter, which we call
AdS giant gravitons, correspond to D-branes expanded into a
$S^3\subset AdS_5$ and the previous section was devoted to study
of their gauge theory dual operators.

We denote by $(\tau,\sigma_1,\sigma_2,\sigma_3)$ the world-volume
coordinates of an AdS giant. The D3-brane equations of motion are
solved by choosing the embedding space-time coordinates to be
given by
\begin{eqnarray}
 && t=\tau\, , \qquad \rho=\rho_0\, ,  \qquad \theta=0\, ,
 \qquad \psi={\tau}\,\nn\\
 && \varphi=\sigma_1\, ,  \qquad  \eta=\sigma_2\, ,
 \qquad \xi = \sigma_3\, . \label{para}
\end{eqnarray}
Independently of the position $\rho_0$ of the giant, the angular
velocity   is $\dot \psi =1$. Thus, its center of mass, located at
$\rho=0$, moves along a null trajectory. Nevertheless,  each
element of the giant travels in a time-like orbit.

The radius of the spherical giant gravitons and their angular
momentum $p$ along the $\psi$ direction  inside the $S^5$ are
given in terms of the radial AdS coordinate $\rho$ by
\begin{eqnarray}
r &=& R\sinh\rho_0\, , \label{r1}
\\
p &=& N \sinh^2\rho_0\, \label{r2} .
\end{eqnarray}
From this equations one gets that $r^2=2 \sqrt{\pi
g_s}\alpha'p/{\sqrt N}$. The DBI action, of which giant gravitons
are D3-branes solutions, is then a valid approximation  if
$p>>\sqrt N$.

Weakly excited strings with a large angular momentum, i.e. those
traveling along almost null trajectories on any spacetime, can be
approximated as moving on an effective pp-wave geometry
\cite{BMN}. We will be interested in open strings having large
angular momentum along $\psi$, the Penrose limit can be understood
as a large $N$ limit, with the angular momentum $p$ growing
proportional to $N$. By looking at (\ref{r1}) and (\ref{r2}) one
realizes that the value $\rho_0$ must be kept constant and
therefore the radius of the giant diverges. The open string will
be effectively attached to a flat D3-brane in a pp-wave
background\footnote{An almost null open string trajectory attached
to an AdS giant requires moving along the $\psi$ direction inside
the $S^5$ and a direction parallel to the giant (see
(\ref{null})). So, the present situation is somehow similar to the
case of non-maximal giants in the $S^5$ \cite{bcv,bcv2} and some
exchange of angular momentum between the D-brane and the open
string is expected.}.

For definiteness, let us consider a trajectory along $\psi$ and
$\eta$, keeping $\varphi=0$. A null trajectory requires
\begin{equation}
R^2(-\cosh^2{\rho_0} {\dot t}^2 +
\sinh^2 \rho_0  {\dot {\eta}}^2 + {\dot \psi}^2)=0 \, .
\label{null}
\end{equation}
Since we want the trajectory to be contained in the D-brane
world-volume, the parametrization (\ref{para}) ${\dot t}=1$ and
${\dot \psi}=1$ necessarily implies $\dot{\eta}= \pm 1$. This
means that the coordinates $t$, $\psi$ and $\eta$ are equally
rated along this null trajectory. This null trajectory is a null
geodesic of $AdS_5\times S^5$ and we use it to take a Penrose
limit (see appendix \ref{ppwave}). Looking at the open string
spectrum on the pp-wave background (\ref{en3}), it is natural to
expect for the first excited states of (\ref{hamiltonian}), the
following eigenvalues in the limit of large $L$ but
$\lambda(\alpha^2-1)^2/L^2$ fixed and small,
\begin{equation}
\label{guess}
E_n \approx \frac{\lambda \pi^2
(\alpha^2-1)^2 n^2 }{L^2}\,.
\end{equation}
However, at the moment we do not know how to diagonalize the
(\ref{hamiltonian}) and make an explicit comparison with
(\ref{guess}). What we do know is the groundstate (\ref{ground}).
Computing its mean occupation number is very elucidative,
\begin{eqnarray}
\label{groundoc}
\frac{1}{L}
\langle\Psi_0 | \hat{n} | \Psi_0\rangle =   \frac{1
}{\alpha^2-1} = \frac{N}{p}\,.
\end{eqnarray}
This computation gives, in average, the number of $Z$ fields in
the word appended over the number of covariant derivatives. The
former ones carry the R-charge identified in the dual description
with angular momentum of the string along $\psi$. The covariant
derivatives carry the spin charge identified in the dual
description with angular momentum of the string along $\eta$.

Now, we can use the null trajectory (\ref{null}) to compute  the
ratio of angular momentum components $J_\psi$ and $J_\eta$ of a
massless particle traveling along it. The ratio of the angular
momenta in both angular directions is
\begin{equation}
\frac{J_\psi}{J_\eta}= \frac{1}{\sinh^2\rho_0}=
\frac{N}{p}\,,
\label{ratio}
\end{equation}
where $p$ is the angular momentum of the giant defined in
(\ref{r2}). This result coincides exactly with (\ref{groundoc}).
Thus, the ground state (\ref{ground}) corresponds to the point
like string. This enforces our interpretation of the first
excitations of the Hamiltonian (\ref{hamiltonian}) as excitation
modes of the open string in the pp-wave background.

\subsection{Semiclassical Open Strings}
\label{som}

Let us now consider a long open string ending on the AdS giant
graviton, that feels the full $AdS_5 \times S^5$ background. One
can see that in the large angular momentum limit the system is
well described by its classical action \cite{Tseytlinrev}.

We concentrate on those coordinates subject to Dirichlet boundary
conditions.  We will eliminate the invariance of
reparametrizations of the world-sheet by fixing a uniform gauge,
instead of the conformal gauge \cite{krurt}. This particular gauge
turns out to be the appropriate one according to the particular
labeling of the operators we used in the dual gauge theory.

We closely parallel the gauge fixing done in \cite{bcv} and
\cite{bcv2} for open strings ending on sphere giant gravitons. Let
us begin by expressing the bosonic Polyakov action in terms of
phase space variables. To do that, we use the conjugate momenta
\begin{equation}
p_\mu= -G_{\mu\nu}(A\partial_0 x^\nu + B \partial_1 x^\nu)\, ,
\end{equation}
where $A=\sqrt{-g} g^{00}$, $B=\sqrt{-g} g^{01}$ and $g^{ab}$ is
the worldsheet metric. The Polyakov action then takes the form
\begin{eqnarray}
\label{paction} S_p = \frac{1}{2\pi \alpha'} \int d\tau \int_0^\pi
{d\sigma} {\cal L}\;,
\end{eqnarray}
where,
\begin{eqnarray}
{\cal L} &=&
-\frac{1}{2}\sqrt{-g}g^{ab} G_{\mu\nu}\partial_a X^\mu \partial_b X^\nu
\nn\\
&=& p_\mu \partial_0 x^\mu + \frac{1}{2} A^{-1} \left[
G^{\mu \nu} p_\mu p_\nu + G_{\mu \nu} \partial_1 x^\mu
\partial_1 x^\nu \right] + B A^{-1} p_\mu \partial_1 x^\mu\;.
\end{eqnarray}
$A$ and $B$ can be thought as
Lagrange multipliers implementing the constraints
\begin{eqnarray}
\label{c1}
G^{\mu \nu} p_\mu p_\nu + G_{\mu \nu} \partial_1 x^\mu
\partial_1 x^\nu & =& 0\, ,\\
\label{c2}
p_\mu \partial_1 x^\mu & =& 0\, .
\end{eqnarray}
We will consider an open string moving with the giant graviton and
along it. In terms of the global coordinates chosen in (\ref{global})
the string will evolve only on $(t,\rho,\eta,\psi)$, i.e. the string
is propagating on an $AdS_3\times S^1 \subset AdS_5\times S^5$,
\begin{equation}
\label{global2}
ds^2 = R^2(-\cosh ^2 \rho dt^2+d\rho^2+\sinh^2\rho d{\eta}^2+d\psi^2) \, .
\end{equation}
In these coordinates the position of the giant is given by
\begin{equation}
\rho = \arg\cosh\sqrt{1+\frac{p}{N}}\,,\qquad
\psi = t\,.
\end{equation}
Now, we change to a coordinate system in which the giant gravitons is
static
\begin{equation}
r = \cosh\rho\,,\qquad
\phi = \psi-t\, .
\end{equation}
We make this election of coordinates guided by the boundary
conditions (\ref{bcr}) and (\ref{bcfi}). The metric
(\ref{global2}) becomes
\begin{equation}
\label{global3}
ds^2 = R^2(-(r^2-1) dt^2+2dt d\phi + d\phi^2
+\frac{dr^2}{r^2-1}+(r^2-1)d{\eta}^2) \, .
\end{equation}
Calling $L$ the total angular momentum of the string along $\eta$
\begin{equation}
L  = \frac{1}{2\pi \alpha'} \int_0^\pi {d\sigma} p_\eta\,.
\end{equation}
As originally done in \cite{AF}, we choose a gauge in which
$p_\eta$ is homogeneously distributed along the string, i.e. it is
independent of $\sigma$. Moreover, we take $\tau$ to be coincident
with the global time $t$,
\begin{eqnarray}
\label{gauge}
t = \tau \,,  \qquad p_\eta = {2 \alpha' L}\,.
\end{eqnarray}
The election of distributing $p_\eta$ homogeneously is also
inspired in the field theory analysis of the previous section. The
spacing in the bosonic lattice was given by the covariant
derivatives. So, the index $l$ running from 1 to $L$ in
(\ref{hamiltonian}) counts uniformly covariant derivatives. Then,
in the continuum limit, the action (\ref{sa2}) has the spin charge
associated to the covariant derivative uniformly distributed in
the variable $\sigma$. This is why we expect the appearance of an
action similar to (\ref{sa2}) by fixing a gauge where $p_\eta$ is
constant. On the contrary, if one worked with the labeling of the
words as $sl(2)$ spin chains, the discrete index of the chain
would count uniformly scalar fields $Z$. In that case, the
semiclassical action obtained projecting with $sl(2)$ coherent
states, would be similar to a Polyakov action in a gauge where the
momentum $p_\psi$ is homogeneously distributed
\cite{Stefanski,Bel}.

The implementation of constraints (\ref{c1}) and (\ref{c2}) leads to
a lagrangian of the form
\begin{eqnarray}
\label{Lagmom}
{\cal L} =V^i p_i  - \sqrt{M^{ij}p_ip_j + M} \, ,
\end{eqnarray}
where the indices $i,j=\phi,r$. The coefficients $V^i$, $M^{ij}$
and $M$ depend on the coordinates and their derivatives,
\begin{eqnarray}
V^{\phi} &=& \dot{\phi}+1\;,\nn\\
V^{r} &=& \dot{r}\;,\\
M^{\phi\phi} \!
&=&\! r^2\left(1+\frac{R^4}{4 \alpha'^2 L^2}(r^2-1){\phi'}^2\right)\;,\nn\\
M^{\phi r}  \! &=&  \!M^{r \phi}
=r^2(r^2-1) \frac{R^4}{4 \alpha'^2 L^2}{r'}\phi'\;,\nn\\
M^{r r} \! &=&\!
r^2(r^2-1) \left( 1+ \frac{R^4}{4 \alpha'^2 L^2}{r'}^2\right)\;,
\\
M &=& 4\alpha'^2 {L}^2\frac{r^2}{r^2-1} +
R^4 r^2 {\phi'}^2 + R^4\frac{r^2}{ r^2-1}{r'}^2\;.
\end{eqnarray}
As usual, dots and primes denote derivatives with respect to $\tau$
and $\sigma$ respectively. According to our conventions,
$R^4/\alpha'^2=\lambda/8\pi^2$. Also, recall that the coordinates
$r$ and $\phi$ satisfy Dirichlet boundary conditions,
\begin{eqnarray}
\label{bc1}
r|_{\sigma = 0, \pi} &=&  \sqrt{1 + \frac{p}{N}}\;, \\
\phi|_{\sigma = 0, \pi} &=& {\rm const.}
\label{bc2}
\end{eqnarray}
Variations of the momenta $p_r$ and $p_\phi$ give rise to algebraic
equations of motion that  can be used to solve the momenta,
\begin{equation}
p_i = M_{ij}V^j\ \sqrt{\frac{M}{1-M_{kl}V^kV^l}}\, ,
\end{equation}
where $M_{ij}$ is the inverse of $M^{ij}$. With these expressions
for the momenta, the Lagrangian can be written as
\begin{eqnarray}
\label{Lfinal}
{\cal L} = -\sqrt{(1-M_{ij}V^iV^j)M} \, ,
\end{eqnarray}
Now, we assume that time derivatives are small. More precisely, we
consider $\partial_0 x^\mu \sim \lambda/{L}^2$ and take this
parameter $\lambda/L^2\ll1$ \cite{kru}. For later convenience, we
rescale $\sigma \rightarrow \sigma/\pi$ and obtain, to lowest
order in the expansion on this small parameter, the following
action
\begin{eqnarray}
\label{stringaction}
S \approx -L \int dt \int_0^1 d\sigma
\left(\frac{\dot{\phi}}{ r^2 - 1} - \frac{\lambda}{L^2} (r'^2 +
r^2 \phi'^2) + {\cal O}\left(\frac{\lambda^2}{L^4} \right)\right)\;.
\end{eqnarray}
Notice the factor $L$ in front of the action, which  $L \rightarrow
\infty$ in the limit we are considering. Then, it can play the r\^ole
of the inverse of the Planck constant in a semiclassical limit.

Remarkably, (\ref{stringaction}) and boundary conditions
(\ref{bc1})-(\ref{bc2}) coincide with the semiclassical
(\ref{sa2}) action for the lattice Hamiltonian and boundary
conditions (\ref{bcr})-(\ref{bcfi}). Therefore, it is natural to
identify coherent states parameters $u$ and $\phi$ with global
coordinates $1/\cosh \rho$ and $\psi-t$ respectively. Moreover, in
this interpretation of the semiclassical coherent state action as
the sigma model action corresponding to an open string, different
labelings of the operators can be accounted by different ways of
fixing the world-sheet reparametrization invariance of the
Polyakov action.

\section{Discussion}
\label{discu}

We have studied excited AdS giant gravitons and their  gauge
theory dual operators. We started  describing the gauge invariant
operators associated to open strings ending on AdS giant
gravitons. To study their scale dimensions spectrum, we made use
of the dilatation operator computed in \cite{Beisert1} restricted
to a $sl(2)$ sub-sector of the super-conformal group. The mixing
matrix of anomalous dimensions, at the one-loop approximation,
corresponded to the Hamiltonian of an open $sl(2)$ spin chain. The
Hamiltonian included terms  mixing  spin chains of different
lengths and at that point we introduced a labeling for the
operators that enabled us to interpret the mixing matrix of
anomalous dimensions as the Hamiltonian of a bosonic lattice. In
analogy with the case of non-maximal sphere giants \cite{bcv}, the
variability in the spin chain length was translated into a
variable total number of bosons occupying the lattice.
Interestingly, the Hamiltonian turned out to be non-quadratic in
the lowering and raising operators and included non-nearest
neighbor interactions. In spite of that, we showed that lattices
with a large number of sites were effectively described by a local
non-linear sigma-model action.

We were not able to compute the complete spectrum of the bosonic
Hamiltonian. However, by a semiclassical analysis, similar to that
of \cite{bcv2}, we showed the existence of continuous bands. In
fact, the variability of the total occupation number and the
existence of continuous bands  prevented one to solve the problem
in terms of a standard Bethe Ansatz. Interestingly, these
continuous bands allow states with a monotonically growing number
of bosons. The duals of such states correspond to open strings
increasing monotonically their angular momentum at the expense of
the giant. The analogy offered in \cite{bcv2} of an open string
with accelerating end points also works here. The instability is
due to the fact that for long enough strings, the tension will not
support the weight due to the centrifugal force.

We have seen that the ratio between the $Z$ fields  and covariant
derivatives ${\cal D}$ mean values for the ground state of the
Hamiltonian (\ref{hamiltonian}) coincides exactly with the ratio
between the angular momentum in the $\psi$ and $\eta$ directions
for an unexcited open string spinning with and along the giant.
This gives support to the proposed dual operators (\ref{go}). In
section \ref{strings} we have presented a semiclassical limit for
the Polyakov action of open strings ending on AdS giant gravitons.
We chose  coordinates in which the giant is static, and fixed a
gauge in which the string angular momentum along AdS, $p_\eta$, is
taken large and uniformly distributed along the string. In this
gauge and in a rigid string limit, the Polyakov action and its
Dirichlet boundary conditions\footnote{We only consider string
excitations along the directions subject to Dirichlet boundary
conditions.} coincided exactly with the non-linear sigma-model
action and boundary conditions of the semiclassical description of
the mixing matrix of anomalous dimensions. The agreement between
these semiclassical actions is intuitively correct within  the
context of the AdS/CFT correspondence. Firstly, the string angular
momentum $p_\eta$ taken large is the geometrical counterpart of
the a large number of sites in the Hamiltonian. Secondly, the
gauge choice that distributes $p_\eta$ uniformly is natural, since
in the continuum limit the number of sites is uniformly rated by a
continuous variable $\sigma$.

One of the motivations for studying AdS giant gravitons was that,
in contrast to sphere giant gravitons, there is no upper bound for
their angular momentum. In this regard, it is particularly
interesting to consider the limit $\alpha \to\infty$ of the
Hamiltonian (\ref{hamiltonian}). In that limit the Hamiltonian can
be diagonalized perturbatively and the terms proportional to
$\alpha^2$ in (\ref{hamiltonian}) serve to define a simple
unperturbed Hamiltonian which, in our base, turns out to be
diagonal. A one boson state in the $k^{\rm th}$ site is an
eigenstate with eigenvalue
\begin{equation}
E_k^0 = \lambda\alpha^2\frac{(L+1)}{k(L+1-k)}\,.
\end{equation}
The first excitations of the unperturbed Hamiltonian are then of
order $\lambda\alpha^2/L$. Although a perturbative treatment might
be a fairly valid approximation for the large $\alpha$ limit of
(\ref{hamiltonian}), it is clear that the one-loop approximation
is no longer correct when taking $\lambda\to\infty$, as in the BMN
limit. Thus, it would be inappropriate to compare the large
$\alpha$ limit of the one-loop Hamiltonian, with any string theory
result.

As a final comment, we would like to point out that the bosonic
labeling developed in section \ref{dilatation} might be useful for
studying other setups. Recently in \cite{samuel}, string bits in
condensates of BPS configurations \cite{BBPS}  and their relation
to giant magnons \cite{HM}, were conveniently characterized using
a similar bosonic labeling for single trace operators in $su(2)$
and $su(3)$ sub-sectors. In terms of the coordinates and gauge
choices appropriate for the bosonic labeling, the string bits or
magnons  are simply depicted as straight lines. We speculate that
the bosonic labeling developed here might be useful to explore
similar magnon excitations in the $sl(2)$ sub-sector.

\section*{Acknowledgements}

We would like to thank D.Berenstein for reading and commenting the
manuscript. G.A.S. would like to thank the Physics group at CECS for
financial support and warm hospitality. D.H.C. work was supported by
Fondecyt grant 3060009 and G.A.S. work by CONICET grant PIP 6160.
Institutional grants to CECS of the Millennium Science Initiative,
Fundaci\'on Andes, and support by Empresas CMPC are also
acknowledged.

\appendix
\section{Conventions and some combinatorial properties}
\label{proper}

We work with double line notation for the gauge theory. All fields
are in the adjoint representation of the gauge group, e.g.
$Z^i_{\;j}=Z^a (T^a)^i_{\;j}$, with $T^a$ the generators of $U(N)$
in the fundamental representation normalized as $\mathrm{tr}(T^a
T^b)=\delta^{ab}$. The key identity is
\begin{equation}
 (T^a)^i_{\;j} (T^a)^k_{\;l}=\delta^i_l\delta^k_j
\end{equation}
~

\noindent We list below some properties of the totally symmetric
tensor that are useful for the calculations of AdS giant gravitons.
The totally symmetric tensor of rank $p$ is defined as
\begin{eqnarray}
S^{i_1\cdots i_p}_{j_1\cdots j_p}\equiv
\sum_{\sigma}\delta^{i_1}_{j_{\sigma(1)}}\cdots\delta^{i_p}_{j_{\sigma(p)}}~,
\label{sym}
\end{eqnarray}
the summation being over all possible permutations $\sigma$. In
(\ref{sym}),  $p$ can be any integer and $i_1,\cdots, i_p$ and $j_1,
\cdots, j_p$ range from $1$ to $N$.  Simple examples are
\begin{eqnarray}
S^{i}_j&=&\delta^{i}_{j}\nonumber\\
S^{ij}_{kl}&=&\delta^{i}_{k}\delta^{j}_{l}+\delta^{i}_{l}\delta^{j}_{k}
\end{eqnarray}
Some  useful properties of the $S$ tensors are
\begin{eqnarray}
 \label{formula1}
S^{i_1\cdots i_p}_{j_1\cdots j_p}&=&
\sum_{x=1}^{p}\delta^{i_1}_{j_x}
S^{i_2\;\;\;\;\;\;.\;\,.\;\,.\;\;\;\;\;\;\;
i_p}_{j_1\cdots j_{x-1}j_{x+1}\cdots j_p} \\
\label{formula2}
S^{i_1\cdots i_k i_{k+1}\cdots i_p}_{i_1\cdots i_k j_{k+1}
\cdots j_p} &=&
\frac{(N+p-1)!}{(N+p-1-k)!}S^{i_{k+1}\cdots i_p}_{j_{k+1}\cdots j_p} \\
\label{formula3}
S^{i_1\cdots i_k}_{j_1\cdots j_k}S^{j_1\cdots j_p}_{l_1\cdots l_p}
&=&k!S^{i_1\cdots i_k
j_{k+1}\cdots j_p}_{l_1\;\;\;\;\; .\;\, .\;\, .\;\;\;\;\;\; l_p}
\end{eqnarray}
Using the previous relations, it is possible to obtain the
following contractions for two $S$ tensors,
\begin{eqnarray}\
\label{identity}
S^{i_1\cdots i_{p-1}a}_{j_1\cdots j_{p-1}b}
S^{j_1\cdots j_{p-1}c}_{i_1\cdots i_{p-1}d}&=&
\frac{(N+p-1)!(p-1)!}{N!}\left(\delta^a_b\delta^c_d
+ \frac{(p-1)}{(N+1)}S^{ac}_{bd}\right)\\
S^{i_1\cdots i_{p-1}ag}_{j_1\cdots j_{p-1}bh}S^{j_1\cdots
j_{p-1}ce}_{i_1\cdots i_{p-1}df}&=&
\frac{(N+p-1)!(p-2)!}{(N+1)!}\left(S^{ag}_{bh}S^{ce}_{df} +
\frac{(p-2)(p-3)}{(N+2)(N+3)}S^{agce}_{bhdf}\right.\nonumber\\
&&\;\;\;+\left.\frac{(p-2)}{(N+2)}\left(\delta^g_h S^{ace}_{bdf}
+\delta^g_b S^{ace}_{hdf}+
\delta^a_h S^{gce}_{bdf}+\delta^a_b S^{gce}_{hdf}\right)\right)
\end{eqnarray}

\section{Open strings in the pp-wave limit}
\label{ppwave}

In this appendix we take a Penrose limit of the $AdS_5\times S^5$
geometry for the null geodesic:
$\rho=\rho_0,\,t=\psi=\eta=\lambda$ discussed in eqn.
(\ref{null}). The limit is accomplished by defining a linear
diffeomorphism to new coordinates and taking the $R\to\infty$
limit. The coordinate $u$ playing the r\^ole of parameter along
the curve must appear, as discussed in section \ref{strings}, with
the same coefficient in $t$, $\psi$ and $\eta$. The remaining
coefficients for the linear transformation can be fixed by
demanding the metric to be well-defined in the $R\to \infty$
limit.

Consider the following change of coordinates,
\begin{eqnarray}
&& t = u +\frac{v}{R^2\cosh^2\rho_0} \, ,
~ ~ ~ ~ ~ ~ ~ ~ ~ ~ ~ ~ ~ ~ ~ ~ ~ ~ ~ ~ ~ ~ ~ ~ ~ ~ ~  ~ ~ ~ ~ ~
\!\rho= \rho_0+\frac{y}{R} \nn\\
&& \psi = u -\frac{v}{R^2\cosh^2\rho_0} -\tanh\rho_0 \frac{x}{R}  \,
,
~  ~ ~ ~ ~ ~ ~ ~ ~ ~  ~ ~ ~ ~ ~\theta= \frac{r}{R} \nn\\
&& \eta =u -\frac{v}{R^2 \cosh^2\rho_0}+
\frac{x}{R\cosh\rho_0\sinh\rho_0}\, , ~  ~ ~ ~ ~ ~  \varphi=
\frac{z}{R\sinh\rho_0}\, . \label{change}
\end{eqnarray}
keeping $\xi$ and $\Omega'_3$ variables unchanged.
After taking $R\to \infty$, the metric (\ref{global}) becomes,
\begin{equation}
ds^2 = -4dudv + 4 y dudx - (r^2+z^2)du^2 +dx^2+dy^2+ dz^2 + z^2
d\xi^2 + dr^2 +r^2d{\Omega_3'}^2 \, . \label{ppomega}
\end{equation}
In terms of cartesian coordinates
\begin{eqnarray}
&& z_1 = z\sin\xi\, , ~ ~ ~ ~ ~ ~ ~ ~ ~ ~ ~ ~ ~ ~ z_4
= r \sin\varphi'\cos\xi'\, \nn\\
&& z_2 = z\cos\xi\, , ~ ~ ~ ~ ~ ~ ~ ~ ~ ~ ~ ~ ~ ~ z_5
= r \cos\varphi'\sin\eta'\, , \nn\\
&& z_3= r \sin\varphi'\sin\xi'\, ,  ~ ~ ~ ~ ~ ~ z_6
= r \cos\varphi'\cos\eta'\, ,
\end{eqnarray}
the metric (\ref{ppomega}) takes the form
\begin{equation}
ds^2 = -4dudv + 4 y dudx - \sum_a z_a^2 du^2 +dx^2+dy^2 + \sum_a
dz_a^2 . \label{ppw}
\end{equation}
This is the well known maximally supersymmetric pp-wave of type IIB
supergravity \cite{Blau} displayed in unusual ``magnetic"
coordinates. This can be explicitly seen by an appropriate change of
variables \cite{Miche,Bertolini}.

The AdS giants attached strings that we consider have two different
angular momenta: one related to the spinning along the $\psi$
direction of the $S^5$ called $J_\psi$ in (\ref{ratio}) and a second
one $J_\eta$ along the $\eta$ direction of the $S^3 \subset AdS_5$
(see eqns.(\ref{global})-(\ref{omegas})). For weakly excited
strings, the quantization on the pp-wave geometry (\ref{ppw}) is a
good approximation. We focus on the bosonic sector of the
corresponding superstring action. In particular, on the excitation
modes of coordinates $x$ and $y$ in (\ref{ppw}) which satisfy
Dirichlet boundary conditions.

The metric (\ref{ppw}) coincides with the one considered in
\cite{bcv2}, so we can then borrow the complete analysis of
oscillation modes and the canonical quantization done in  it. The
light-cone Hamiltonian expressed in terms of the oscillator
operators $\beta_m, \tilde\beta_m, \beta_m^\dagger$ and
$\tilde\beta_m^\dagger$ of the $x$ and $y$ coordinates is
\begin{eqnarray}
H_{lc}^{xy}\!&=&\!\frac{1}{8{\alpha'}^2 p^u}
\int_0^{2\pi\alpha'p^u}\!\!\!\!d\sigma
\left(\dot x^2 +\dot y^2 + {x'}^2+{y'}^2\right)
\nn\\
&=& \frac{1}{2{\alpha'} p^u}\sum_{n>0}
\left(\omega_n^-\tilde\beta_n^\dagger\tilde\beta_n
+\omega_n^+\beta_n^\dagger\beta_n\right)\,,
\end{eqnarray}
where $\omega_n^\pm = \sqrt{(2 \alpha' p^u)^2+n^2} \pm 2 \alpha' p^u$.
Thus, the energy of each excitation is
\begin{equation}
\label{en}
\tilde E_n = \sqrt{1+\frac{n^2}{(2 \alpha' p^u)^2}}-1\, ,
 \qquad E_n = \sqrt{1+\frac{n^2}{(2 \alpha' p^u)^2}}+1\;.
\end{equation}
To  illuminate  this result, we  express the light-cone charges
\begin{eqnarray}
H_{lc} \!&=&\! -p_u = i\frac{\partial}{\partial u}\, ,\\
p^u\!&=&\! -\frac{1}{2}p_v=\frac{i}{2}\frac{\partial}{\partial v}\,.
\end{eqnarray}
in terms of the conserved charges $ \Delta$, $J_\psi$ and $J_\eta$
corresponding to the original global coordinates $t,\psi$ and
$\eta$. The change of coordinates (\ref{change}) gives,
\begin{eqnarray}
H_{lc} &=& i\left(\frac{\partial}{\partial t}+
\frac{\partial}{\partial \psi}
+\frac{\partial}{\partial \eta}\right)= \Delta - J_\psi -J_\eta\, ,\\
p^u &=& \frac{i}{2 R^2 \cosh^2\rho_0}\left(\frac{\partial}{\partial t}
- \frac{\partial}{\partial \psi}
-\frac{\partial}{\partial \eta}\right)
=\frac{\Delta + J_\psi +J_\eta}{2R^2  \cosh^2\rho_0}\,.
\end{eqnarray}
To make contact with the gauge theory calculations, we denote
\begin{eqnarray}
J_\eta &=& L\, ,\\
J_\psi &=& \frac{L}{\sinh^2\rho_0} =\frac{N}{p}L\,.
\end{eqnarray}
$L$ denotes, on the gauge theory side, the total number of covariant
derivatives present in the word $W$ (see eqn. (\ref{L})). The sum of
angular momenta takes the form
\begin{equation}
J_\eta+J_\psi = {L}\left(1+\frac{N}{p}\right)
= {L}{\frac{\alpha^2}{\alpha^2-1}}\,,
\end{equation}
where   $\alpha$ is given by (\ref{alfa}).  To get a finite
light-cone energy we required that $\Delta\simeq \alpha^2
L/(1-\alpha^2)$. Therefore,
\begin{equation}
p^u\simeq \frac{L}{R^2(\alpha^2-1)}
= \frac{L}{\sqrt{8\lambda}\pi(\alpha^2-1)}\,.
\end{equation}
Here the relations  $R^4 = 4\pi g_s N \alpha'^2$ and $\lambda=g_s
N/2\pi$ have been used. Finally, the string excitation energies in
terms of gauge theory parameters are
\begin{equation}
\label{en2} \tilde E_n = \sqrt{1+\frac{2\lambda \pi^2
(\alpha^2-1)^2 n^2 }{L^2}}-1\ ,
 \qquad E_n = \sqrt{1+\frac{2\lambda \pi^2
(\alpha^2-1)^2 n^2 }{L^2}}+1\;.
\end{equation}
 When  considering the BMN limit: $\lambda/L^2$ fixed and small
 as $L\to\infty$, the expansion of  the square roots gives,
\begin{equation}
\label{en3}
\tilde E_n \approx \frac{\lambda \pi^2
(\alpha^2-1)^2 n^2 }{L^2}\, ,  \qquad E_n \approx 2+
\frac{\lambda \pi^2 (\alpha^2-1)^2 n^2 }{L^2}\;.
\end{equation}


\end{document}